\documentclass[aps,prl,twocolumn,showpacs,floatfix,superscriptaddress]{revtex4-1}

\usepackage{graphicx,color}
\usepackage{times}
\usepackage{amsmath,amssymb}
\usepackage{amsthm}
\usepackage{mathrsfs}
\usepackage{multirow}
\usepackage{natbib}
\usepackage[colorlinks,citecolor=red]{hyperref}

\newcommand{\re}{\operatorname{Re}}

\newcommand{\Eqnref}[1]{Equation~(\ref{#1})}
\newcommand{\figref}[1]{Fig.~\ref{#1}}

\begin{document}

\title{Quantum Resistor-Capacitor Circuit with Majorana Fermion Modes in Chiral
  Topological Superconductor}

\author{Minchul Lee}
\affiliation{Department of Applied Physics, College of Applied Science, Kyung Hee University, Yongin 446-701, Korea}
\author{Mahn-Soo Choi}
\affiliation{Department of Physics, Korea University, Seoul 136-701, Korea}
\date{\today}

\begin{abstract}
  We investigate the mesoscopic resistor-capacitor circuit consisting of a
  quantum dot coupled to spatially separated Majorana fermion modes in a
  chiral topological superconductor.
  We find substantially enhanced relaxation resistance due to the
  nature of Majorana fermions, which are their own anti-particles and
  composed of particle and hole excitations in the same abundance.
  Further, if only a single Majorana mode is involved, the
  zero-frequency relaxation resistance is completely suppressed due to a
  destructive interference. As a result, the Majorana mode opens an exotic
  dissipative channel on a superconductor which is typically regarded as
  dissipationless due to its finite superconducting gap.
\end{abstract}

\maketitle

As electronic circuit is miniaturized on the nanometer scale, quantum
coherence takes effect and transport properties get fundamentally
different. For a ballistic conductor, Ohm's law breaks down and the conductance
is quantized to multiples of $R_Q\equiv h/e^2$ \cite{vanWees88a,Wharam88a},
where $h$ is the Planck constant and $e$ is the electron charge. For a small
resistor-capacitor circuit, the charge relaxation resistance is also quantized
to $R_Q/2$, irrespective of the transmission properties
\cite{Buttiker1993sep,Buttiker1993jun}, as demonstrated in an experiment on a
quantum dot (QD) coupled to quantum Hall (QH) edge channel
\cite{Gabelli2006jul}.
The quantization is technically ascribed to the fermi-liquid nature of the
system
\cite{Nigg2006nov,LeeMC2011may,Mora2010jun,Filippone2011oct,KhimH2013mar},
where the relaxation of particle-hole (p-h) pairs due to charge
fluctuations at the cavity is the culprit for the dissipation.
It is tempting and indeed customary \cite{Buttiker1999apr} to interpret the quantized value
as the contact resistance at a single interface (hence a half of the two-terminal contact resistance $R_Q$).

\begin{figure}[!b]
  \centering
  \includegraphics[width=4.5cm]{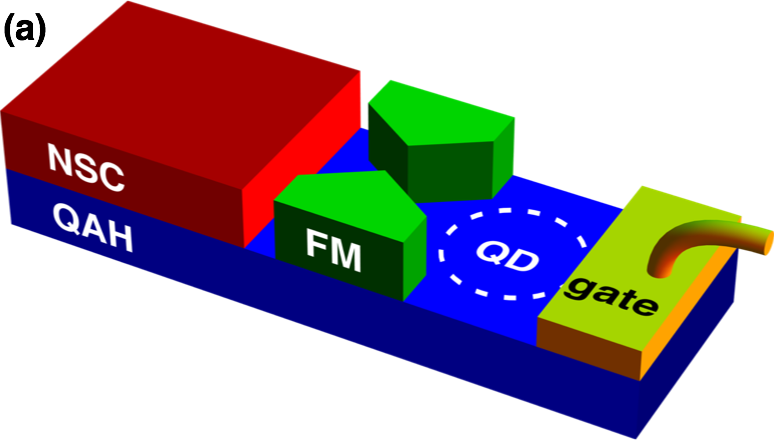}%
  \includegraphics[width=4cm]{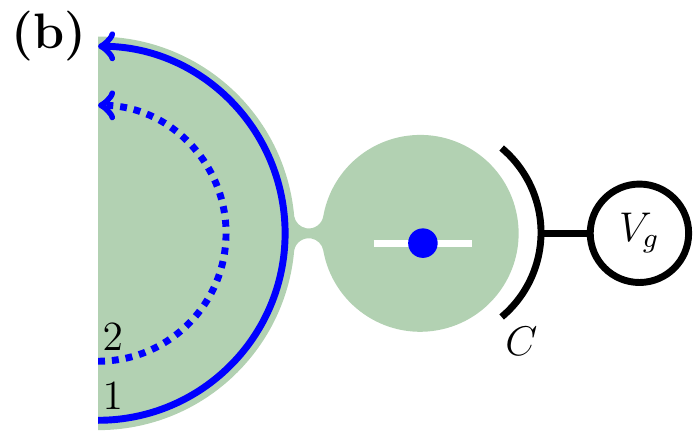}
  \caption{(Color online) (a) A possible realization of quantum capacitor with spatially separated Majorana fermion modes.
    (b) A schematic of the coupling between two
    spatially separated Majorana edge modes ($j=1,2$) and the energy level
    localized on the QD.
  }
  \label{fig:1}
\end{figure}

Here we show that when the circuit involves Majorana fermions, which
are casually regarded as half-fermions, the quantum resistance defies such an
interpretation. Specifically, we examine a QD coupled, with different
strengths, to two \emph{spatially separated} one-dimensional (1D) chiral
Majorana fermion modes; see \figref{fig:1}. In the ultimate limit, a single
Majorana mode is considered. The primary goal is to identify the role of
each Majorana mode in relaxation resistance and compare it to the case
of Dirac fermion mode.

Mathematically, a Dirac fermion can always be decomposed into a pair of
Majorana fermions, but these Majorana fermions usually occupy the same spatial
location. However, the chiral topological superconductor (cTSC) states
\cite{Qi2010nov} enable physical realization of spatially separated 1D Majorana
fermion modes. An example is a quantum anomalous Hall (QAH)
insulator proximity-coupled to a conventional (or \textit{normal}) superconductor (NSC) \cite{Qi2010nov}.
A HgTe quantum well doped with Mn element \cite{Liu08b} and a Bi$_2$Te$_3$ thin
film doped with Cr element \cite{Yu10b,Chang2013apr} turn into a QAH
insulator with a chiral Dirac fermion edge mode, i.e., two chiral
Majorana edge modes. When the QAH insulator is coupled to a NSC (see \figref{fig:1}), the proximity-induced pairing potential
pushes one of the two Majorana modes deeper into the bulk, spatially separating it from the
other. As the relative magnitudes of the magnetization and the superconducting
gap vary, the system undergoes topological phase transitions, from QAH
insulator phase to a cTSC phase (hereafter called as the cTSC$_2$ phase) with
two spatially separated Majorana edge modes \cite{endnote:1}, to another cTSC
phase (called as the cTSC$_1$ phase) with a single Majorana edge mode (one mode
having disappeared into the bulk), and finally to a NSC
without any edge channel \cite{Qi2010nov}.

\begin{figure}[!t]
  \centering
  \includegraphics[width=8cm]{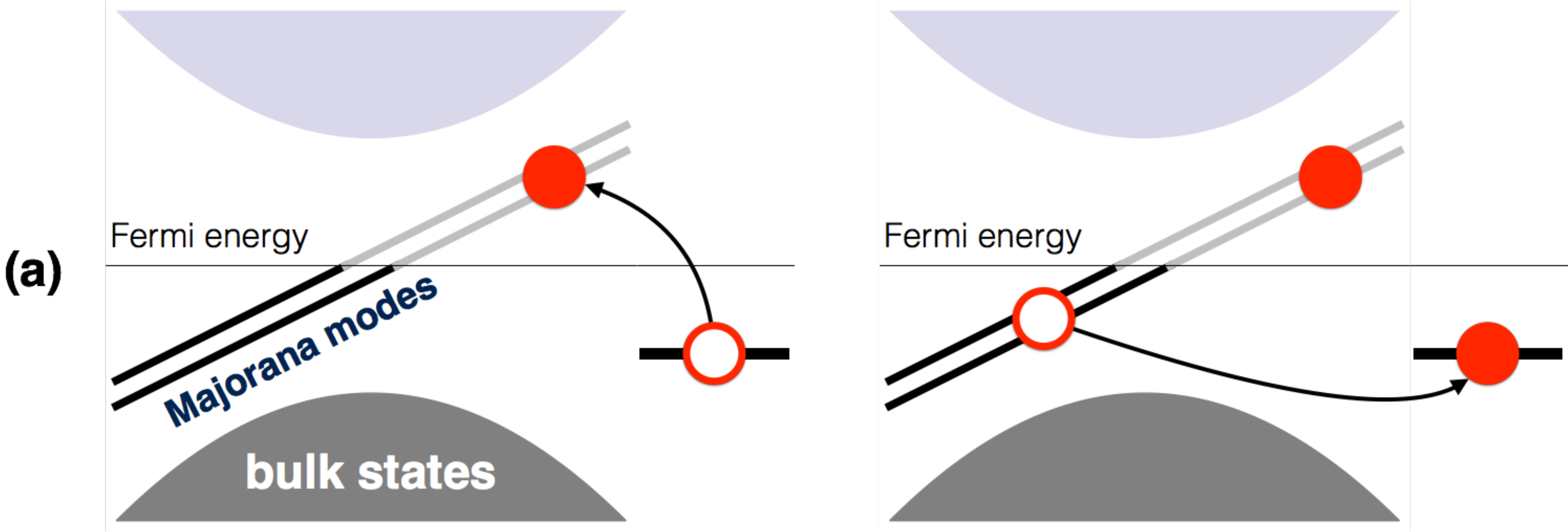}
  \includegraphics[width=8cm]{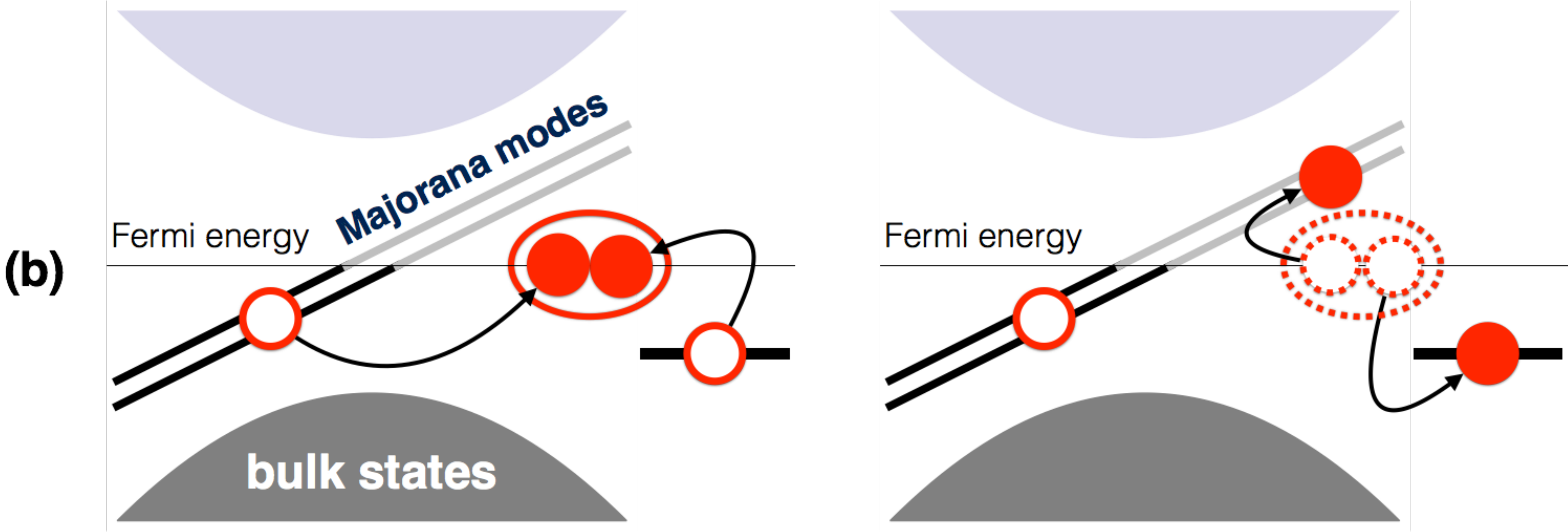}
  \caption{(Color online) Second-order processes to generate a p-h pair in the
    Majorana fermion channel via (a) the charge-conserving process ($\sim
    |t_\text{single}|^2$) and (b) the pairing process ($\sim
    |t_\text{pair}|^2$) when the QD is initially occupied.
    The degenerate Majorana modes are artificially
    split here as a guide for the eye.}
  \label{fig:3}
\end{figure}

Once the system enters the cTSC phases (either cTSC$_1$ or cTSC$_2$), we find
that the low-frequency relaxation resistance is no longer pinned at $R_Q/2$
and strongly depends on the transmission properties.
Especially, as the QD level approaches the resonance, the zero- or
finite-frequency resistance is substantially enhanced,
suggesting that Majorana modes boost the p-h pair generation and are
highly dissipative. It contrasts with the gapped superconductor case, in which
the resistance is suppressed for frequencies $\omega$ smaller than the
gap.  For the cTSC$_1$ phase with only a single Majorana edge mode, on the
other hand, we find that the low-frequency relaxation resistance vanishes in
the $\omega\to0$ limit as for the fully gapped superconductor.  The vanishing
resistance is attributed to the exact cancellation between p-h pair generation
processes in charge-conserving and pairing channels, as will be discussed later
(see \figref{fig:3}).  These exotic behaviors are distinguished
from those for normal superconductors or Dirac fermion channels.
This casts an intriguing question about the role of the Majorana fermions in
relaxation resistance and offers {another method to probe the
Majorana fermions}.

\paragraph{Model.---}
Focusing on the low-energy physics inside the bulk gap, one can describe a cTSC
with two chiral Majorana modes
\begin{align}
  \label{eq:HMajorana}
  H_\text{Majorana}
  = \sum_{j=1,2}\sum_{k>0} \epsilon_k\gamma_{-k,j} \gamma_{k,j},
\end{align}
where $\gamma_{k,j}=\gamma_{-k,j}^\dag$ are chiral Majorana fermion operators,
$\epsilon_k=\hbar v k$ is their energy, and $v$ is the propagation velocity of
the Majorana edge modes.
In the cTSC$_1$ phase, we regard the mode $j=2$ disappearing into the bulk.

The QD can be formed by depositing ferromagnetic insulators (FMs), which turns
the underneath region into the trivially insulating state (I). A proper
placement of FMs deforms and localizes the QAH edge states to form
a QD; see \figref{fig:1}(a). Since the localized state in the QD originates from
the spin-polarized QAH edge state, it is described as a single spinless
level $\epsilon_d$:
\begin{align}
  \label{eq:HQD}
  H_{\rm QD} = \left\{\epsilon_d + e[U(t) - V_g(t)]\right\} n_d.
\end{align}
Here $n_d = d^\dag d$ is the occupancy operator, and the ac voltage $V_g(t)$
upon the gate coupled to the QD via a geometrical capacitance $C$ induces the
polarization charge on the dot and eventually the
internal potential $U(t)$. The latter is determined self-consistently under the
charge conservation condition.

The coupling of the QD level to the chiral Majorana edge modes ($j=1,2$)
takes a tunneling model \cite{Zitko2011may}
\begin{align}
  \label{eq:HC}
  H_\text{tun}
  =
  \sum_k
  \left[t_1 d^\dag\gamma_{k,1} + i t_2 d^\dag\gamma_{k,2} + {\rm (h.c.)}\right].
\end{align}
Here, for simplicity, we have assumed wide bands and neglected the momentum
dependence of the tunneling amplitudes $t_j$ between the Majorana mode $j$ and
the QD level. In this limit, the coupling is conveniently described by the
hybridization parameters
\begin{math}
\Gamma_j \equiv |t_j|^2/\hbar{v}
\end{math}
(\begin{math}
\sim 0.4\text{--}4\operatorname{\mu eV}
\end{math}\cite{Gabelli2006jul})
and $\Gamma_\pm\equiv(\Gamma_2\pm\Gamma_1)/2$.
In general $\Gamma_1\geq\Gamma_2$ due to their spatially separated localizations;
in particular, $\Gamma_2=0$ in the
cTSC$_1$ phase and $\Gamma_1=\Gamma_2$ only in the QAH phase.
Note that our model ignores the bulk states of the reservoir and
$\Gamma_1$ and $\omega$ should be
sufficiently smaller than the bulk gap; it is
inadequate when the system is too close to the cTSC$_2$-cTSC$_1$ transition
point, where the bulk gap is small. $\Gamma_2$ may vanish well before the transition point due to the exponential suppression with distance of the tunneling.

\begin{figure}[!t]
  \centering
  \includegraphics[width=60mm]{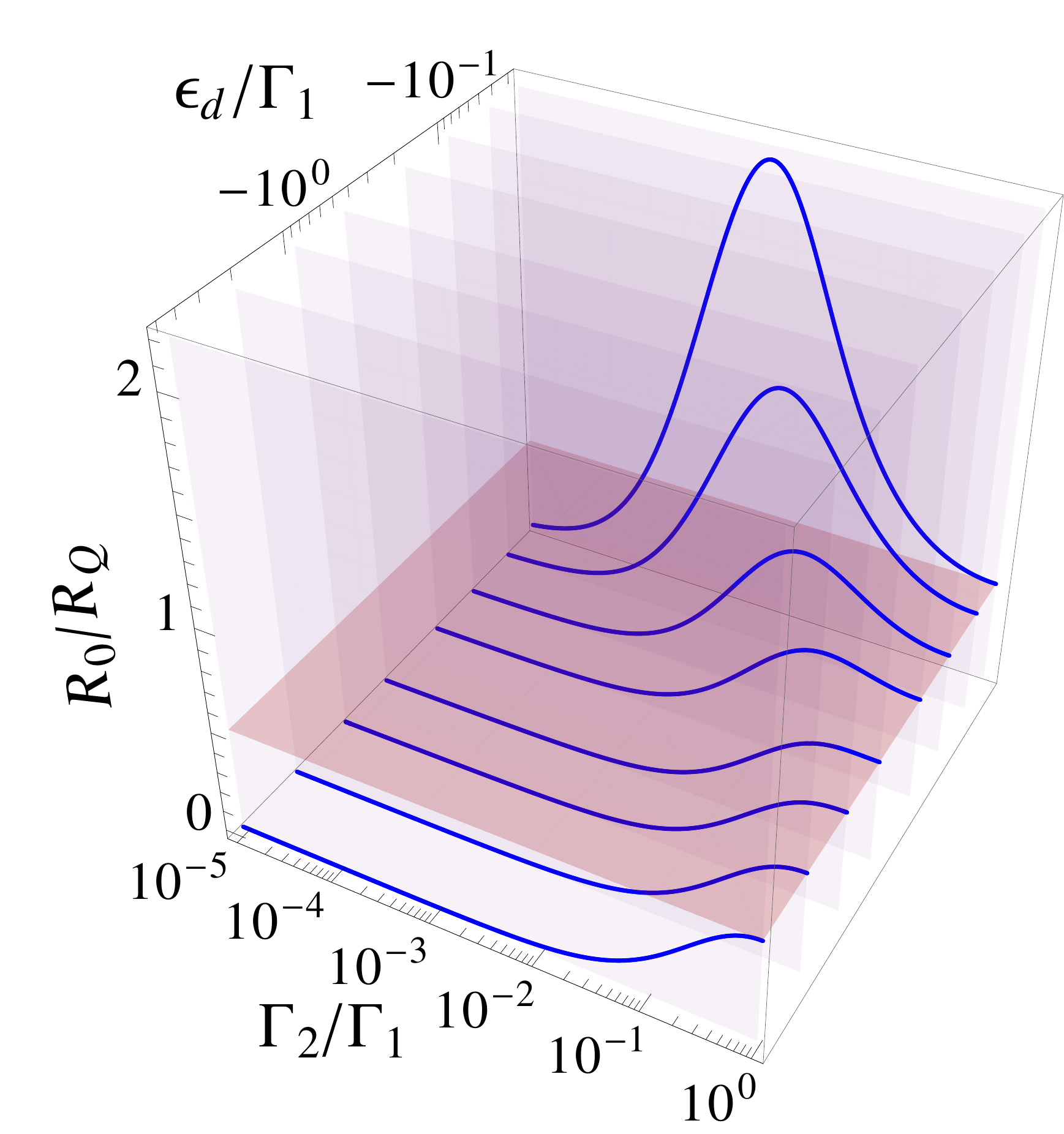}
  \caption{(Color online) Zero-frequency relaxation resistance $R_0 =
    R_q(\omega\to0)$ as a function of $\epsilon_d$ and $\Gamma_2$. For
    comparison, a shaded horizontal surface is also drawn at the quantized value
    $R_Q/2$.}
  \label{fig:2}
\end{figure}

\paragraph{Relaxation Resistance.---} We calculate the ac current $I(t)$
between the reservoir of Majorana modes and the QD, or equivalently the
displacement current between the top gate and the QD, using the self-consistent
mean-field approach in the linear-response regime
\cite{Buttiker1993sep,Buttiker1993jun,Nigg2006nov,LeeMC2011may,Mora2010jun,Filippone2011oct,KhimH2013mar,Jauho1994aug}
(see \cite[Sec.~II.A,B]{endnote:2}).
The
relaxation resistance $R_q(\omega)$ is then obtained
from its relation to the admittance $g(\omega)$, $1/g(\omega) = R_q(\omega) +
i/\omega C_q(\omega)$, where $C_q(\omega)$ is the quantum correction to the
capacitance.
At zero temperature the admittance allows for a closed-form expression and
reads as (hereafter we set $\hbar=k_B=1$)
\begin{multline}
  \label{eq:admittance}
  g(\omega)
  =
  \frac{1}{R_Q}
  \sum_{\mu=\pm}
  \Bigg\{
  \frac{\Gamma_-^2}{\varepsilon(\omega+\mu\varepsilon)}
  \ln\frac{\Gamma_++i\varepsilon}{\Gamma_+-i\varepsilon}
  \\\mbox{}
  +
  \left[
    \frac{\Gamma_-^2}{\varepsilon(\varepsilon+\mu\omega)}
    +
    \frac{\Gamma_+}{\Gamma_++i\omega}
  \right]
  \ln\frac{\Gamma_+ + i(2\omega+\mu\varepsilon)}%
  {\Gamma_+ + i\mu\varepsilon}
  \Bigg\}
\end{multline}
with $\varepsilon \equiv \sqrt{4\epsilon_d^2 - \Gamma_-^2}$.
\Eqnref{eq:admittance} is the main result of this work. We now discuss its
physical implications.

\paragraph{Zero-frequency resistance at zero temperature.---}
Let us first focus on the zero-frequency limit ($\omega \ll
\epsilon_d^2/\Gamma_1$) of the resistance, $R_0\equiv R_q(\omega\to0)$; see \figref{fig:2} and \cite[Sec.~II.C]{endnote:2}.
In the QAH phase, where the two Majorana modes equally contribute
($\Gamma_1=\Gamma_2$), the resistance restores the quantized value, $R_0 =
R_Q/2$, as expected because the two Majorana modes in the QAH phase are
equivalent to a single Dirac fermion mode.
%

As the system evolves into the cTSC$_2$ phase ($0<\Gamma_2<\Gamma_1$), $R_0$
does not only deviate from the quantized value but also depends on the ratio
$\Gamma_2/\Gamma_1$ and the QD level $\epsilon_d$, as shown in \figref{fig:2}.
When the dot level is far from the resonance ($|\epsilon_d|\gg\Gamma_1$), the
zero-frequency resistance,
\begin{math}
R_0 \approx
\left(R_Q/2\right)
\left[4(\Gamma_2/\Gamma_1)/(1+\Gamma_2/\Gamma_1)^2\right],
\end{math}
depends only and monotonically on the ratio $\Gamma_2/\Gamma_1$; see the curve
for large values of $|\epsilon_d|$ in \figref{fig:2}.
When the dot level resonates with the Fermi level
($|\epsilon_d|\ll\Gamma_1$), it now depends non-monotonically on
$\Gamma_2/\Gamma_1$ with the maximum at $\Gamma_2 \approx \Gamma_m \equiv
4\epsilon_d^2/\Gamma_1$ of the height
$\sim[4\gamma_m\ln\gamma_m]^{-1}$ with $\gamma_m\equiv\Gamma_m/\Gamma_1$.
%
In short, unlike the Dirac fermion case, $R_0$ for the reservoir of Majorana
modes strongly depends on the properties of the tunneling barrier and the QD,
and thus defies the simple interpretation \cite{Buttiker1999apr} of it
as a half of the two-terminal contact resistance.

The zero-frequency relaxation resistance in the cTSC$_1$ phase with a single
Majorana mode ($\Gamma_2=0$) is even more interesting and exotic: it vanishes
exactly, $R_0=0$, irrespective of $\Gamma_1, \Gamma_2$ and $\epsilon_d$, although there is no excitation energy gap \cite[Sec.~II.C]{endnote:2}.
To understand it, we introduce chiral Dirac fermion
operators $c_k\equiv(\gamma_{k,1}+i\gamma_{k,2})/\sqrt{2}$ composed of the two
Majorana fermions, in terms of which the Hamiltonians~(\ref{eq:HMajorana}) and
(\ref{eq:HC}),
respectively, are
rewritten as
\begin{align}
  H_\text{Majorana}
  & = \sum_k \epsilon_k c_k^\dag c_k
  \\
  \label{eq:Htun:DF}
  H_\text{tun}
  & =
  \sum_k
  \left[
    t_\text{single}d^\dag c_k + t_\text{pair}d^\dag c_k^\dag
    + {\rm (h.c.)}
  \right]
\end{align}
with $t_\text{single/pair} \equiv (t_{1}\pm t_{2})/\sqrt{2}$.  This
form~(\ref{eq:Htun:DF}) immediately suggests two distinctive types of processes,
as illustrated in \figref{fig:3}:
One is charge-conserving type from the $t_\text{single}$-term, in which the
p-h pair is excited via the electron tunneling in and out of the QD [\figref{fig:3}(a)]. This type of processes alone would give rise to $R_0 =
R_Q/2$ \cite{LeeMC2011may,Mora2010jun,Filippone2011oct,KhimH2013mar}. The other
is pairing type involving the $t_\text{pair}$-term which accompanies the
creation and destruction of a Cooper pair in the bulk [\figref{fig:3}(b)]. This type is missing in the QAH phase, where $t_1=t_2$.
When the QD is initially occupied, the charge-conserving (pairing) process creates the
particle (hole) first. Hence the p-h
pair amplitudes of the two processes are opposite in sign  (at all orders) due to the fermion ordering. When $\Gamma_2 = 0$
($t_\text{single}=t_\text{pair}$), both types are the same in magnitude so as
to cancel out each other exactly. This cancellation and the subsequent
vanishing resistance are hallmarks of the relaxation via the Majorana
modes. Note, however, that this cancellation is exact only for
$\Gamma_2=0$ and for p-h pairs with vanishingly small energy ($\omega\to0$
limit). At finite $\omega$, as shown below, the intermediate
virtual states are different for two processes so that the
cancellation is not perfect.

\begin{figure}[!t]
  \centering
  \includegraphics[width=4cm]{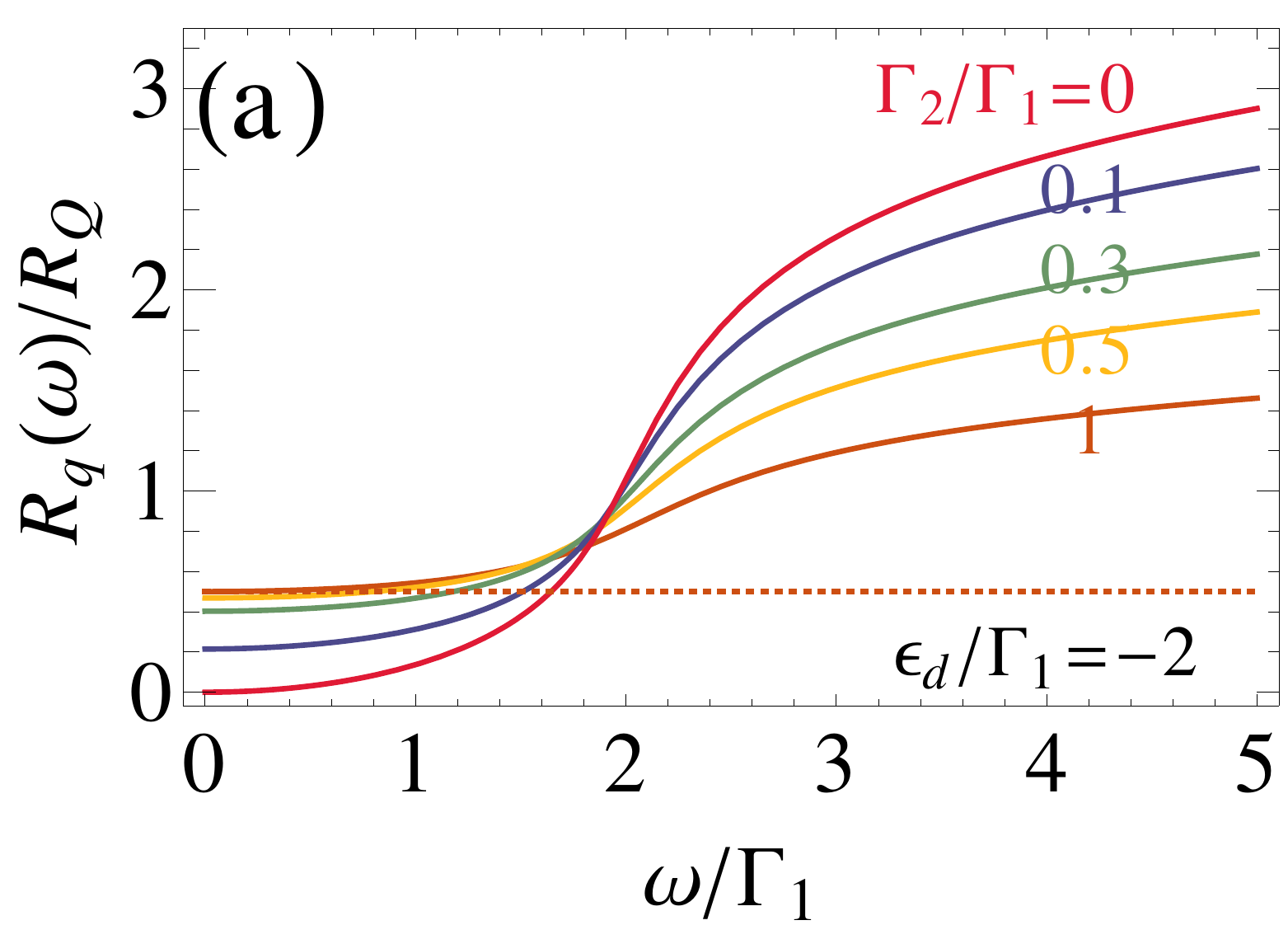}
  \includegraphics[width=4cm]{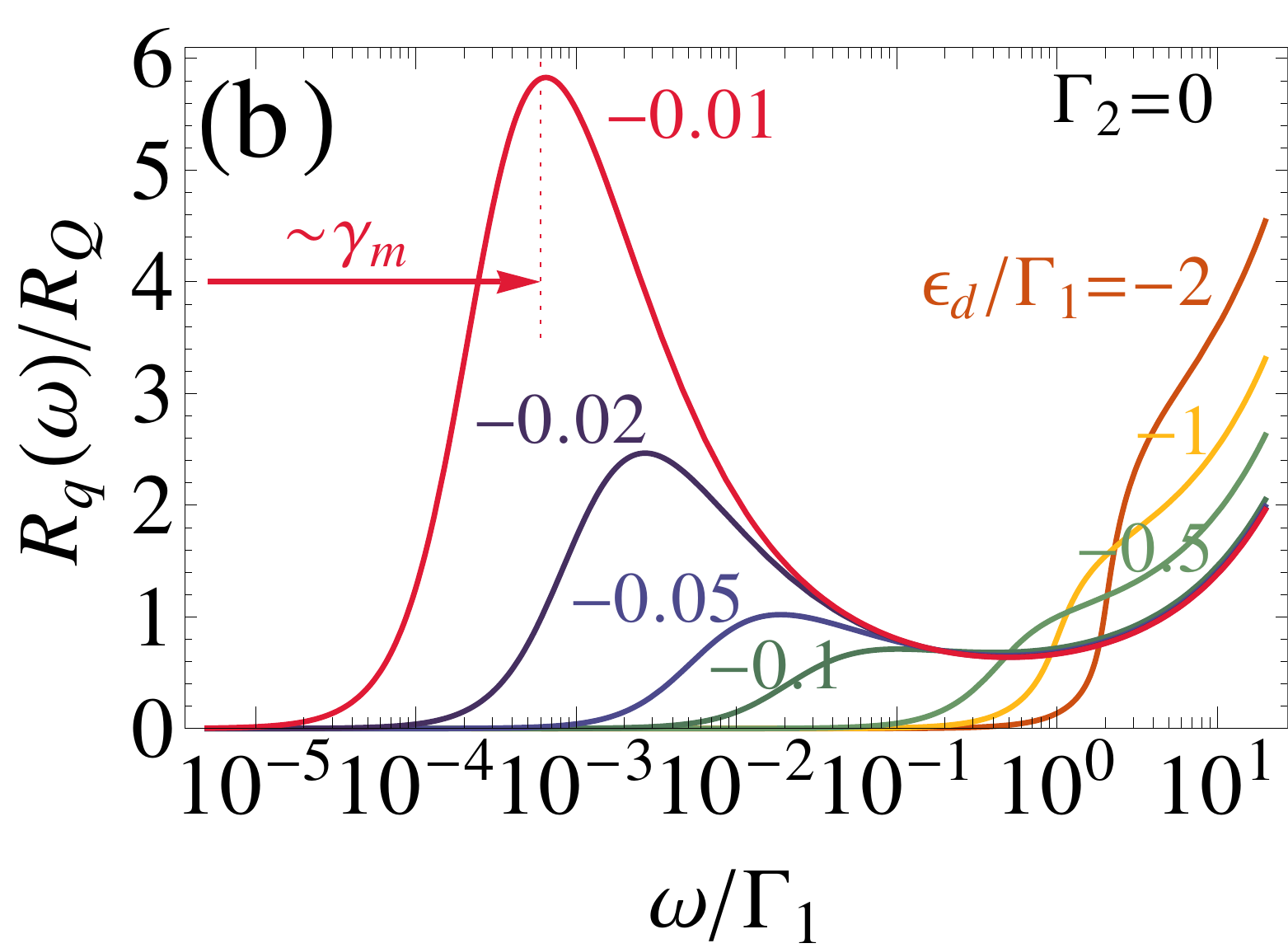} \par
  \includegraphics[width=4cm]{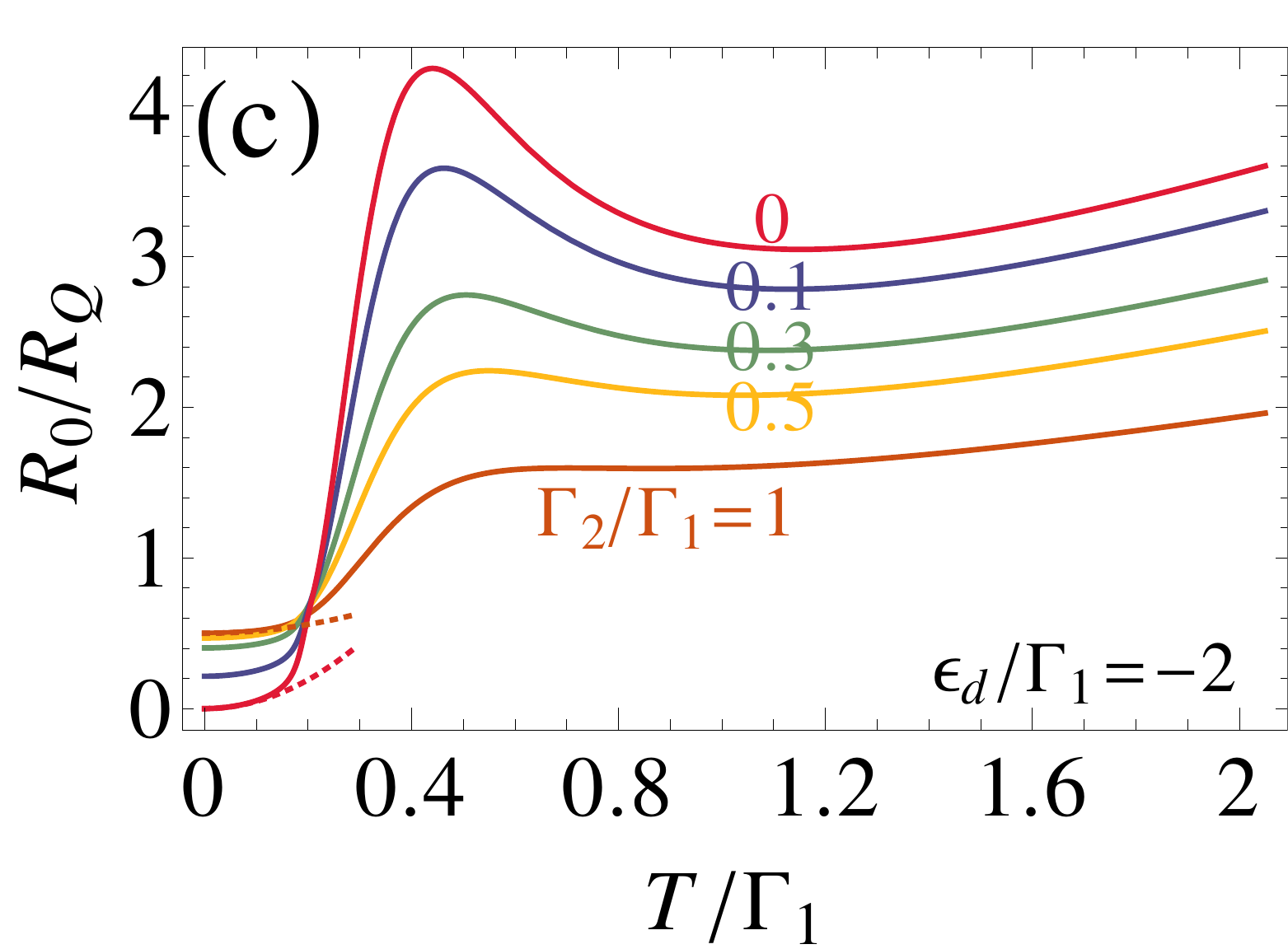}
  \includegraphics[width=4cm]{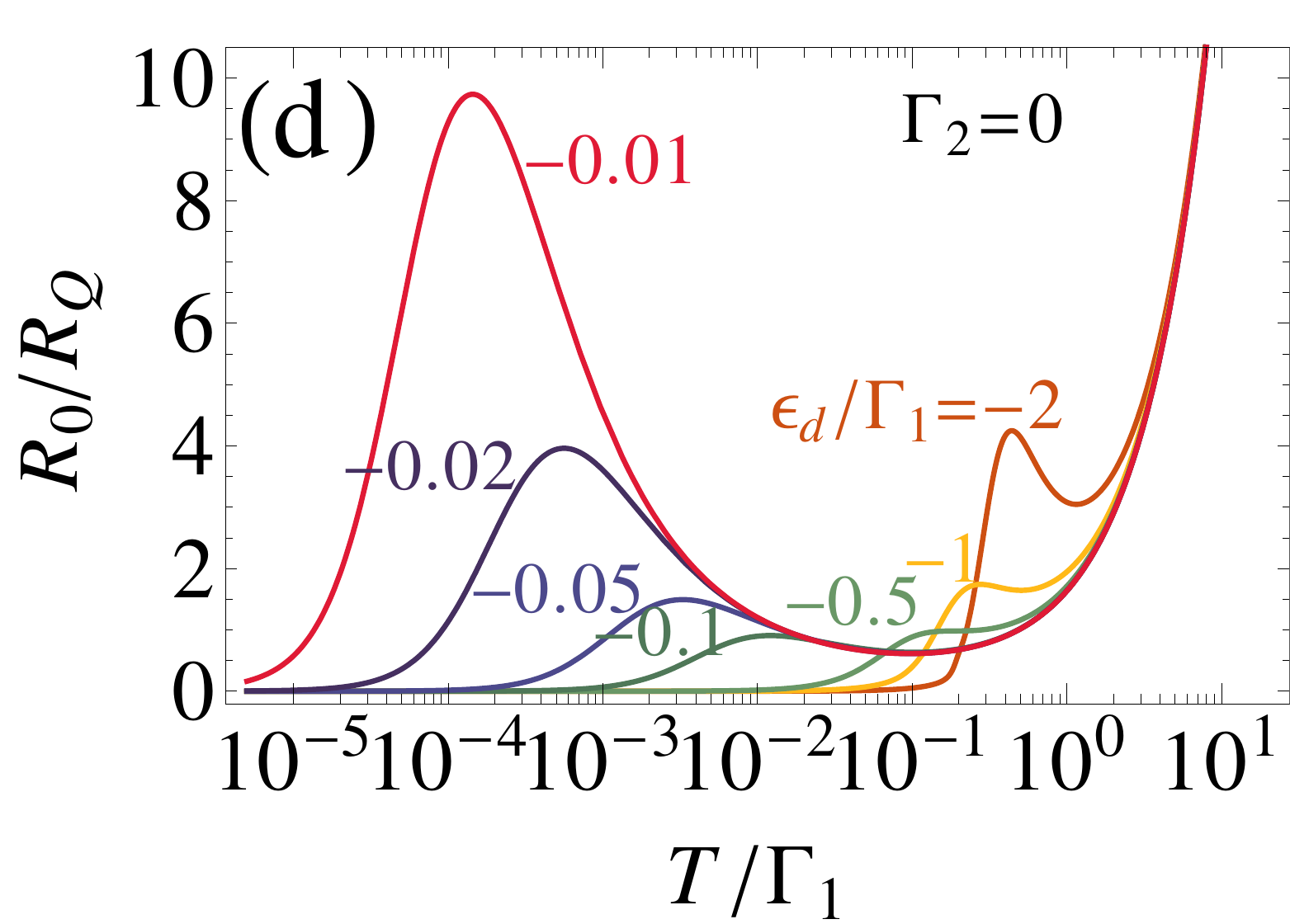}
  \caption{(Color online) (a,b) Zero-temperature resistance
    $R_q(\omega)=R_q(-\omega)$ as a function of frequency $\omega$. (c,d)
    Zero-frequency resistance $R_0$ as a function of temperature
    $T$. $\epsilon_d/\Gamma_1=-2$ in (a,c), and $\Gamma_2=0$ in (b,d). The
    arrow in (b) indicates the approximate peak position $\sim\gamma_m$ for
    $\Gamma_2=0$ and $\epsilon_d/\Gamma_1 = -0.01$. The dotted lines in (c)
    correspond to the low-temperature asymptotes.}
  \label{fig:4}
\end{figure}

\paragraph{Finite-frequency resistance at zero temperature.---}
We
find that the vanishing or
enhancement of the resistance discussed above becomes even more pronounced at
finite frequencies.
While the finite-frequency resistance $R_q(\omega)$ is zero for conventional
superconducting reservoir {or even for Andreev bound states localized at yet propagating along its edge (\emph{Andreev edge modes}) due to their finite gap \cite[Sec.~III.D]{endnote:2}, it grows
with $|\omega|$ for the gapless Dirac fermion reservoir since the spectral
density of p-h excitations grows with energy \cite{LeeMC2011may}.
In the QAH phase, $R_q(\omega)$ grows slowly and monotonically with $\omega$ [\figref{fig:4}(a)].
Entering the cTSC phase, however, $R_q(\omega)$ becomes highly
non-monotonic, forming an ever narrower dip
at $\omega=0$ and peaks at
$|\omega|\sim\Gamma_m$ [\figref{fig:4}(a,b)]. The dip width is the order of $\Gamma_m$ for $\Gamma_2 \approx 0$.

Let us examine the cTSC$_1$ phase ($\Gamma_2=0$)
in three different regimes: the (i) off-resonance
($|\epsilon_d|\gg\Gamma_1$), (ii) near-resonance ($0<|\epsilon_d|\ll\Gamma_1$),
and (iii) exact-resonance ($\epsilon_d=0$) regime.  (i) In the off-resonance
regime, the resistance grows like
\begin{math}
R_q(\omega) \approx
(R_Q/3)\left(\omega/\epsilon_d\right)^2
\end{math}
for small frequencies ($\omega\ll\Gamma_m$), and keeps growing monotonically
for higher frequencies.  Note that in this limit the resistance is independent
of the barrier transmission. (ii) In the near-resonance regime,
\begin{math}
R_q(\omega) \approx
\left[R_Q/3\gamma_m(\ln\gamma_m)^2\right]\left(\omega/\Gamma_m\right)^2
\end{math}
for $\omega\ll\Gamma_m$, and it shows sharp peaks at
$\omega\approx\pm\Gamma_m$ [\figref{fig:4}(b)].  The quadratic growth
and dip-peak structure around $\omega=0$
cast stark contrasts with the superconducting reservoir {or the case with Andreev edge modes, in the latter case the peak is either pinned at the gap energy or
  linearly dependent on the QD level \cite[Sec.~III.E]{endnote:2}.
(iii) Even more dramatic contrast appears at the exact resonance.  In this case, the two peaks at $\omega\approx\pm\Gamma_m$ in
\figref{fig:4}(b) merge together, filling up the dip at $\omega=0$ (i.e., the
dip width is zero).  As a result, the resistance
\begin{math}
R_q(\omega) \approx
\pi R_Q/4(2|\omega|/\Gamma_1)(\ln(2|\omega|/\Gamma_1))^2
\end{math}
diverges as $\omega\to0$.
In summary, the finite-frequency relaxation resistance is genuinely enhanced
for $\Gamma_2\ll\Gamma_1$ near resonance.

\paragraph{All-Majorana representation.---} To understand the enhancement of
the resistance near resonance, it is instructive to describe the dot level in
terms of the language of Majorana fermions as well \cite[Sec.~II.D]{endnote:2}. We define
two Majorana operators $\gamma_{d,j}$ ($j=1,2$) by $\gamma_{d,1} =
(d-d^\dag)/\sqrt2i$ and $\gamma_{d,2} = (d+d^\dag)/\sqrt2$.
Unlike
the Majorana fermions on the edge modes of the cTSC these dot Majorana fermions
are rather mathematical as they occupy the same spatial location.  The QD and
coupling Hamiltonians (\ref{eq:HQD}) and (\ref{eq:HC}), respectively, read as
\begin{subequations}
  \begin{align}
    H_{\rm QD}
    & = i\epsilon_d \gamma_{d,2} \gamma_{d,1}
    \\
    H_\text{tun}
    & = \sum_k i (t_2 \gamma_{d,2} \gamma_{k,2} - t_1 \gamma_{d,1} \gamma_{k,1}).
  \end{align}
\end{subequations}
In this expression, $\epsilon_d$ becomes the coupling between the two dot
Majorana fermions. The two Majorana edge modes in the reservoir are coupled
\emph{indirectly} via the coupling between two Majorana fermions on the dot,
being completely decoupled at the resonance ($\epsilon_d=0$). However, it does
not mean that their contributions are independent [see
Eq.~(\ref{eq:g:resonance}) below], because the charge is always composed of two
Majorana fermions.
At $\epsilon_d=0$, the real part of admittance, representing the dissipation,
is expressed as
\begin{align}
  \label{eq:g:resonance}
  \re[g(\omega)]
  =
  \frac{2\pi^2}{R_Q}\omega
  \int_0^\omega  d\omega' \rho_1(\omega-\omega') \rho_2(\omega')
\end{align}
where $\rho_i(\omega')$ is the density of states for $\gamma_{d,i}$, which is Lorentzian, centered at $\omega'=0$ and with
width $\Gamma_i$.
If both $\rho_1$ and $\rho_2$ are finite, then $\re[g(\omega)]\sim\omega^2$ for $\omega\to0$ so that $R_0$ is finite; recall $g(\omega) \approx -i\omega C_q + \omega^2
C_q^2 R_0$. However, as $\Gamma_2\to0$, $\rho_2(\omega')$ becomes sharper and eventually $\rho_2(\omega')=\delta(\omega')$ at $\Gamma_2=0$,
so that $\re[g(\omega)] \sim \omega$,
i.e., $R_0\propto 1/\omega$ as seen above. In short, the resistance enhancement  at resonance is
attributed to a decoupled dot Majorana with abundant density of states
near zero energy;
the dot
\emph{electron} is coupled equally to the particle and hole components of the single Majorana
edge mode so that the Majorana nature, leading to
proliferation of p-h pairs, is highly pronounced.
A single local Majorana fermion coupled to a chiral Majorana line has appeared in a different context, essentially a two-channel Kondo model, and a similar divergence $R_q\sim 1/\omega(\log\omega)^2$ has been observed~\cite{Mora13a}. This suggests that the exotic behaviors of our system may be a non-Fermi liquid feature.

For $\epsilon_d\ne0$, $\gamma_{d,1}$ and $\gamma_{d,2}$ are coupled and
interfere each other, causing anti-Fano-like destructive interference:
The broadening of $\gamma_{d,2}$ is
\begin{math}
\sim \Gamma_m
\end{math}
and the destructive interference leads to
a dip in $\rho_1(\omega)$ of width $\sim\Gamma_m$ [\figref{fig:4}]. This is another explanation, now based on the interference between Majorana
fermions, of the vanishing low-frequency resistance discussed before.

\paragraph{Decoherence Effects.---} We remark that all the features discussed
so far --- the vanishing low-frequency resistance and the divergence of the
resistance \emph{at resonance} --- occur only when the full coherence is
maintained.
In the presence of decoherence, the resistance would deviate from those
coherent values. For example, when the dot is subject to random background
charge fluctuations, which are the most common decoherence source on QDs, it
leads to the fluctuation in $\epsilon_d$. In effect, it pushes the system away
from the resonance and the resistance does not diverge. Another
indication of decoherence effects can be seen in the finite-temperature
effect discussed below.

\paragraph{Finite-temperature effect.} Typically $R_q$ increases with
temperature $T$ since the thermal fluctuations promote the generation of p-h
pairs \cite{Nigg2006nov}. For $T \ll \Gamma_m$, the
Sommerfeld expansion \cite[Sec.~II.E]{endnote:2} gives rise to
\begin{math}
R_0 \approx
R_0|_{T=0}
+ R_Q(2\pi^2/3) \left(T/\epsilon_d\right)^2
\left(1 + \Gamma_-^2/\Gamma_+^2\right)
\end{math}
in the off-resonance regime [\figref{fig:4}(c)].
For higher temperatures, however, non-monotonic behavior is
observed for $\Gamma_2 \ll \Gamma_1$ [\figref{fig:4}(d)]: A peak occurs at
$T \sim \Gamma_m$, whose height grows as $|\epsilon_d|$ decreases. The
enhancement of $R_q$ around this particular temperature is related to the peak
structure in zero-temperature $R_q(\omega)$ located at $\omega \sim \Gamma_m$
as shown in \figref{fig:4}(b). In the presence of thermal fluctuations, the
contribution of low-energy p-h pairs that are suppressed due to the destructive
interference between $\gamma_{d,1}$ and $\gamma_{d,2}$ decreases. Instead the
p-h pairs, which have energy $\sim\Gamma_m$ and are not affected by the
cancellation, cause the surge of the resistance. Together with the
non-monotonic frequency dependence of $R_q(\omega)$, the peak structure driven
by the thermal excitations are unique features of dissipation via Majorana
states.

\paragraph{Acknowledgments.}
This work was supported by the NRF grants funded by the Korea government (MSIP)
(Nos. 2011-0030046 and 2011-0012494).

\bibliographystyle{apsrev}
\bibliography{Paper}

\end{document}